\documentclass[prb,twocolumn,showpacs,preprintnumbers,amsmath,amssymb]{revtex4}
\usepackage{graphicx}
\usepackage{epsfig}
\newcommand{\be}{\begin{eqnarray}}
\newcommand{\ee}{\end{eqnarray}}

\newcommand{\bi}{\begin{itemize}}
\newcommand{\ei}{\end{itemize}}

\begin{document}
\title{Surface deformation caused by the Abrikosov vortex lattice}

\author{Pavel Lipavsk\'y$^{1,4}$, Klaus Morawetz$^{2,3}$,
Jan Kol\'a\v cek$^4$ and Ernst Helmut Brandt$^5$}
\affiliation{
$^1$ Faculty of Mathematics and Physics, Charles University,
Ke Karlovu 3, 12116 Prague 2, Czech Republic}
\affiliation{
$^2$Forschungszentrum Rossendorf, PF 51 01 19, 01314 Dresden, Germany}
\affiliation{
$^3$Max Planck Institute for the Physics of Complex
Systems, Noethnitzer Str. 38, 01187 Dresden, Germany}
\affiliation{
$^4$Institute of Physics, Academy of Sciences,
Cukrovarnick\'a 10, 16253 Prague 6, Czech Republic}
\affiliation{
$^5$Max Planck Institute for Metals Research,
         D-70506 Stuttgart, Germany}
\begin{abstract}
In superconductors penetrated by Abrikosov vortices the magnetic
pressure and the inhomogeneous condensate density induce a
deformation of the ionic lattice. We calculate how this
deformation corrugates the surface of a semi-infinite sample. The
effect of the surface dipole is included.
\end{abstract}

\pacs{
74.20.De, 
74.25.Ld, 
74.25.Qt, 
74.81.-g
}
\maketitle
\section{Introduction}
Deformations of the ionic lattice caused by the Abrikosov vortices
have been studied from several aspects. For example, if a vortex
moves, the motion of such deformation demands a motion of ions
which contributes to the inertial mass of the
vortex.\cite{Sim91,DSim92,CK03} In the static case, forces evoked
by vortices create a tension which modifies the total volume of
the sample and which was observed as magnetostriction.\cite{C94}
Moreover, in anisotropic materials the elastic energy caused by
vortices depends on the relative orientation of the Abrikosov
vortex lattice and the crystal lattice.\cite{UZD73,KBMD95}

In the above mentioned studies a vortex was treated as infinitely
long. This idealized geometry essentially simplifies the problem.
Since the system has translational symmetry along the vortex line,
the lattice deformation is purely longitudinal with the
displacement vectors perpendicular to the vortex.\cite{Sim91,C94}

\begin{figure}
\psfig{file=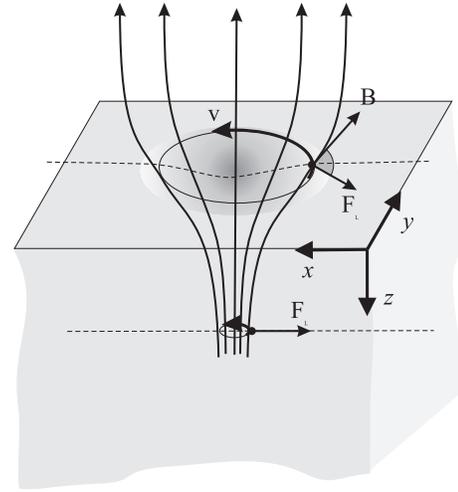,width=6cm}
\caption{Lorentz force acting on the circulating superconducting
current. In the bulk the Lorentz force is parallel to the
surface. Near the surface it is not parallel due to the
magnetic stray field.}
\label{fig_vor}
\end{figure}

As far as we know, nobody has studied deformations near
the surface, where the magnetic flux of the vortex leaves
the superconductor. In the present paper we focus on this problem.
For simplicity we assume that the sample is semi-infinite and the
applied magnetic field is perpendicular to its surface, see
figure~\ref{fig_vor}.

Far from the surface the lattice deformation approaches
its bulk value with displacement vectors perpendicular to
vortices, i.e., parallel to the surface. Let us sketch in
advance which complications one can expect near the surface.

As seen in figure~\ref{fig_vor}, the stretching magnetic
field together with circulating currents result in
a Lorentz force with a component pointing along the vortex.
Electric fields balancing this force act also on the ion lattice
and corrugate the surface.

Beside the Lorentz force, one can imagine two additional
mechanisms leading to the corrugation. First, the specific
volume of the superconducting and normal states differ.
The normal metal in the vortex core is pulled by its
neighborhood to adopt the specific volume of the
superconductor. As mentioned above, deep in the bulk
these stresses cause displacements perpendicular to the
vortex. Close to the surface, however, the same stresses
lead to different deformations as they are partially relaxed
by displacements perpendicular to the surface, i.e., along
the vortex.

Second, the surface dipole
which confines the electrons in the crystal changes under
transition from the normal to the superconducting state.
Accordingly, at the vortex core the surface dipole differs
from other regions of the surface. The force which holds
electrons inward naturally pulls the oppositely charged ions
outwards and also corrugates the surface. The general relation
between the surface dipole and the surface tension has been
established already two decades ago.\cite{KZ86,KZ88} The
contribution of the superconducting condensate has been
discussed only recently.\cite{LMKBS08}

The paper is organized as follows. In sections~II and III
we introduce the basic set of equations for elastic
deformations of an isotropic material near the surface.
The surface deformation due to the Abrikosov vortex lattices
is discussed in section~IV and the numerical solu\-tions for Nb
and YBa$_2$Cu$_3$O$_7$ are presented in section~V. Section~VI with the
conclusions ends the paper.

\section{Elastic deformations in isotropic medium}

In this section we recall the basic definitions needed to
describe the elastic deformations of an isotropic medium. More
details the reader can find in the textbook of Landau and
Lifshitz.\cite{LL75}

\subsection{Tensors of strain and stress}
Deformations are described by atomic displacements
$\bf  u$. We assume that all atoms are identical and
only a single atom occupies the elementary cell.
Since the inter-atomic distance is short compared
to the characteristic scales of deformations,
we treat ${\bf  u}(x,y,z)$ as a function
of continuous coordinates.

The space derivatives of the displacement define the
strain tensor. For small deformations assumed here
its components read
\begin{equation}
u_{xy}={1\over 2}\left({\partial u_x\over\partial y}+
{\partial u_y\over\partial x}\right).
\label{e1}
\end{equation}
The complementary anti-symmetric combination of derivatives
corresponds to the rotation of the rigid body,
therefore it does not contribute to deformations.

The strain creates a stress described by a tensor
$\sigma$. In general $\sigma_{\kappa\tau}=
\sum_{\mu\nu}\Lambda_{\kappa\tau\mu\nu}u_{\mu\nu}$,
where $\Lambda$ is $4^{\rm {th}}$ order tensor of
elastic coefficients. The Greek indices stand for $x,y,z$.
For an isotropic medium this relation simplifies to
\begin{equation}
\sigma_{\kappa\tau}=
K(\nabla\cdot{\bf  u})\,\delta_{\kappa\tau}+
2\mu\left(u_{\kappa\tau}-{1\over 3}
(\nabla\cdot{\bf  u})\,\delta_{\kappa\tau}\right),
\label{e2}
\end{equation}
where the divergence of the displacement vector
\begin{equation}
(\nabla\cdot{\bf  u})=u_{xx}+u_{yy}+u_{zz}
\label{e3}
\end{equation}
is a shorthand notation for the trace of the strain
tensor. It represents the local change of the specific
volume.
The tensor $u_{\kappa\tau}-{1\over 3}
(\nabla\cdot{\bf  u})\,\delta_{\kappa\tau}$ has
zero trace and describes purely shear deformations.
The coefficients $K$ and $\mu$ are called the
bulk and shear modulus, respectively.

\subsection{Stability conditions}

The stress depends on local gradients of the displacements,
i.e., on changes of the bond lengths between neighboring
atoms. Accordingly it represents only contact forces while
long-range forces have to be covered separately. Here we
shall consider a non-contact force due to interaction of
the crystal lattice with superconducting electrons.

The gradient of the stress balances a long-range force
${\bf  F}$ acting on a unitary volume
$\sum_\kappa\nabla_\kappa\sigma_{\kappa\tau}+F_\tau=0$,
where $\nabla_x\equiv{\partial\over\partial x}$ and so on.
For the isotropic material this stability condition
reads\cite{LL75}
\begin{equation}
\left(K+{4\over 3}\mu\right)\nabla(\nabla\cdot{\bf  u})-
\mu\left[\nabla\times\left[\nabla\times{\bf  u}\right]\right]
={\bf  F}.
\label{e4}
\end{equation}

In this paper we consider forces that can be expressed as
a gradient of the potential $U$
\begin{equation}
{\bf  F}=-\nabla U.
\label{e5}
\end{equation}
In our numerical studies below we assume that this potential
is exclusively due to the electrostatic potential $\varphi$
acting on the charge density of the ionic lattice $\rho$, i.e.,
$U=\rho\varphi$.

Alternatively, one can assume an effective potential
$U=2K\alpha_0|\psi|^2/n$ corresponding to the force used
by {\v S}im{\'a}nek and Duan\cite{Sim91,DSim92} and
Coffey\cite{C94}. Here $\alpha_0$ is the relative volume
difference between the superconducting and the normal
states, $K$ is the bulk modulus, $\psi$ is the
Ginzburg-Landau (GL) wave function and $n$ is the
density of pair-able electrons.

These two choices of the potential $U$ yield very similar results.
They are not completely identical, however. For example,
the electrostatic potential known as the Bernoulli potential
covers the Lorentz force, while the effective potential
$U=2K\alpha_0|\psi|^2/n$ does not.\cite{LMKB07} 

\subsection{Surface conditions}

At the surface the balance of forces demands
$\sum_\kappa n_\kappa\sigma_{\kappa\tau}+P_\tau=0$,
where ${\bf  n}$ is a unit vector normal to the
surface and ${\bf  P}$ is the external force on a
unit area of the surface. The normal component
$p=({\bf  n}\cdot{\bf  P})$ is the pressure.
We will not assume tangential surface forces in
this paper.

To simplify boundary conditions we specify the geometry
of our sample. It is a semi-infinite superconductor
in the half-space $z>0$. For the normal vector
${\bf  n}=(0,0,-1)$ and normal surface force
${\bf  P}=(0,0,-p)$ we find three surface
conditions
\begin{equation}
\left(
\begin{array}{ccc@{\ }r}
\sigma_{xz} \\ \sigma_{yz} \\ \sigma_{zz} \\
\end{array}
\right)
=
\left(
\begin{array}{ccc@{\ }r}
0 \\ 0 \\ -p \\
\end{array}
\right).
\label{e6}\\
\end{equation}

Using the linear relation between stress and strain,
the surface conditions (\ref{e6}) are converted into
conditions for the displacement $\bf  u$. For the
isotropic medium from (\ref{e2}) and (\ref{e6}) follows
\begin{eqnarray}
u_{xz}&=&0,
\nonumber\\
u_{yz}&=&0,
\label{e7}\\
\left(K+{4\over 3}\mu\right)u_{zz}+
\left(K-{2\over 3}\mu\right)
\left(u_{xx}+u_{yy}\right)&=&-p.
\nonumber
\end{eqnarray}

The pressure $p$ includes the ambient pressure of coolant which
is homogeneous all over the sample
surface. We ignore it, because it does not contribute to
the corrugation.

We do not want to list all possible forces acting
on the surface. We merely mention that the surface
of the sample feels the surface dipole $\delta\varphi$,
which acts as an effective pressure $p=-\rho\delta\varphi$,
see Ref.~\onlinecite{LMKBS08}. The dipole amplitude
$\delta\varphi=\varphi(0)-\varphi(\infty)$ is the
difference between the electrostatic potential at the
surface and deep in the bulk of the  superconductor.
The surface dipole is conveniently evaluated from the
GL free energy with the help of the Budd-Vannimenus
theorem\cite{BV73,KW96} adopted to
superconductor\cite{LMKMBS04,LMKBS08}.

Now the problem is fully specified. The displacement
$\bf  u$ is driven by the potential $U$ and the effective
pressure $p$. These quantities can be treated within any
approximation the reader prefers. In section~V we will relate 
them to the
electrostatic potential $\varphi$ and the surface
dipole $\delta\varphi$. At the moment we merely 
assume that the potential and the pressure are known 
functions of the GL wave function.
Since the deformation has only a negligible effect on the
GL wave function, the GL wave function obtained from the
ordinary GL theory can be used. We take the GL wave
function as known and focus on the deformation.

\section{Induced and free deformation}
As mentioned in the introduction, the deformation \mbox{near}
the surface has three sources -- magnetic stray field,
expansion or contraction along the vortex, and the
surface dipole. Since the elastic equation (\ref{e4})
and its boundary conditions (\ref{e7}) are linear in
$\bf  u$, it is possible to separate individual
contributions.

As the first step, it is advantageous to separate
the displacement $\bf  u$ into two parts -- induced deformation
${\bf  u}^{\rm i}$ driven by the long range force $\bf  F$
which induces displacements all over the sample,
and the free deformation ${\bf  u}^{\rm f}$.
So we write
\begin{equation}
{\bf  u}={\bf  u}^{\rm i}+{\bf  u}^{\rm f},
\label{e8}
\end{equation}
where the displacement ${\bf  u}^{\rm i}$
is longitudinal
\begin{equation}
[\nabla\times{\bf  u}^{\rm i}]=0.
\label{e9}
\end{equation}
and obeys the equation
\begin{equation}
\left(K+{4\over 3}\mu\right)(\nabla\cdot{\bf  u}^{\rm i})=-U({\bf  r},z).
\label{e10}
\end{equation}
Since the differential equation (\ref{e10}) is of the first
order, this induced displacement is fully specified by the
potential $U$ and the requirement of convergence deep in the sample.

After substitution of (\ref{e8}) into the elastic
equation (\ref{e4}), the stress from the induced displacement
compensates the long range forces and one is left with the equation
for the free displacement
\begin{equation}
\left(K+{4\over 3}\mu\right)\nabla(\nabla\cdot{\bf  u}^{\rm f})-
\mu\left[\nabla\times\left[\nabla\times{\bf  u}^{\rm f}\right]\right]
=0.
\label{e11}
\end{equation}
This is a second-order differential equation and we need to find
the solution which in addition to the convergence deep inside guarantees
also the fulfillment of the surface boundary condition.

\subsection{Laplace equation for the free deformation}
Now we show that it is possible to simplify equation (\ref{e11})
to the Laplace equation. 

Taking the divergence of equation (\ref{e11}) 
we find that
\begin{equation}
\nabla\cdot\left(\nabla^2{\bf  u}^{\rm f}\right)=0,
\label{e11div}
\end{equation}
because the divergence of a rotation is zero. Taking the rotation
of equation (\ref{e11}) we find
\begin{equation}
\left[\nabla\times\left(\nabla^2{\bf  u}^{\rm f}\right)\right]=0,
\label{e11rot}
\end{equation}
because the rotation of a gradient is also zero.
The vector identity $[\nabla\times[\nabla\times{\bf u}^{\rm f}]]=
\nabla(\nabla\cdot{\bf u}^{\rm f})-\nabla^2{\bf u}^{\rm f}$ was
used in the rearrangement.

According to (\ref{e11div}) and (\ref{e11rot}) all the derivatives of
$\nabla^2{\bf  u}^{\rm f}$ are zero, therefore it is a constant.
A finite value of ${\bf  u}^{\rm f}$ for $z\to\infty$
is possible only if this constant is zero.
We thus have
\begin{equation}
\nabla^2{\bf  u}^{\rm f}=0.
\label{e21}
\end{equation}
Being free of material parameters, this Laplace equation is more
convenient than the equation (\ref{e11}).

\subsection{Surface matching}
The surface boundary condition (\ref{e7}) applies to
the total displacement. Substituting (\ref{e8})
into (\ref{e7}) we obtain
\begin{eqnarray}
u^{\rm f}_{xz}&=&-u^{\rm i}_{xz},
\nonumber\\
u^{\rm f}_{yz}&=&-u^{\rm i}_{yz},
\label{e23}\\
\left(K+{4\over 3}\mu\right)u^{\rm f}_{zz}
&+&\left(K-{2\over 3}\mu\right)
\left(u^{\rm f}_{xx}+u^{\rm f}_{yy}\right)
\nonumber\\
=-p-\left(K+{4\over 3}\mu\right)u^{\rm i}_{zz}&-&
\left(K-{2\over 3}\mu\right)
\left(u^{\rm i}_{xx}+u^{\rm i}_{yy}\right).
\nonumber
\end{eqnarray}
These conditions determine gradients of ${\bf  u}^{\rm f}$ at $z=0$.
The induced deformation ${\bf  u}^{\rm i}$ is already known, therefore
we have moved its strain elements to the right hand sides of
the boundary conditions.

By conditions (\ref{e23}) the free displacement is fully specified.
The second boundary condition is the request of convergency,
$u^{\rm f}\to 0$ for $z\to\infty$. Now the set of equations is
complete and we are ready to solve for the corrugation of the surface
around the vortex ends.

\section{Vortex lattice}
If the vortices form an Abrikosov lattice, we can benefit
from its periodic structure. In this case all components
of the displacement can be solved analytically in the
two-dimensional (2D) Fourier representation.

\subsection{Induced deformation}
In the $x$--$y$ plane, the potential is a sum of planar
waves
\begin{equation}
U({\bf r},z)=\sum_{\bf k}
{\rm e}^{i\bf k \bf r}U({\bf k},z),
\label{e24}
\end{equation}
where ${\bf r}\equiv(x,y)$ is a 2D coordinate.
The 2D wave vectors ${\bf k}=(k_x,k_y)$ attain discrete
values given by the density of vortices and the actual
structure of the Abrikosov lattice.

To solve the equation for the displacement requires some
intermediate steps. Since the induced displacement
${\bf  u}^{\rm i}$ is longitudinal, it can be expressed as
the gradient of a scalar function
\begin{equation}
{\bf  u}^{\rm i}=\nabla\chi({\bf r},z)€.
\label{e25}
\end{equation}
In the 2D Fourier representation the equation (\ref{e10})
yields
\begin{equation}
\left(K+{4\over 3}\mu\right)
\left(k^2-{\partial^2\over\partial z^2}\right)\chi
=U({\bf k},z),
\label{e26}
\end{equation}
where $k^2=k_x^2+k_y^2$. It is solved by
\begin{equation}
\chi({\bf k},z)=
{1\over 2k}{1\over K+{4\over 3}\mu}
\int\limits_0^\infty dz'
{\rm e}^{-k|z-z'|}U({\bf k},z').
\label{e27}
\end{equation}

We are interested namely in values at the surface
\begin{equation}
\chi({\bf k},0)=
{1\over 2k}{1\over K+{4\over 3}\mu}
\int\limits_0^\infty dz'
{\rm e}^{-kz'}U({\bf k},z').
\label{e28}
\end{equation}
The $z$-gradient at the surface follows from
(\ref{e27}) as
\begin{equation}
u^{\rm i}_{z}=\left.{\partial\over\partial z}
\chi({\bf k},z)\right|_{z=0}=
k\chi({\bf k},0).
\label{e29}
\end{equation}
The surface value of this displacement contributes to the surface
corrugation.

The surface strain given by the second gradient
follows from the equation (\ref{e26}) as
\begin{equation}
u^{\rm i}_{zz}=\left.{\partial^2\over\partial z^2}
\chi({\bf k},z)\right|_{z=0}=
k^2\chi({\bf k},0)-
{1\over K+{4\over 3}\mu}U({\bf k},0).
\label{e30}
\end{equation}
This surface strain enters the boundary condition,
where it serves as the source of the free deformation.

\subsection{Free deformation}
The free deviation ${\bf  u}^{\rm f}$ has the same periodicity
as the Abrikosov vortex lattice. From the Laplace equation
(\ref{e21}) thus follows
\begin{equation}
{\bf  u}^{\rm f}({\bf k},z)={\bf  u}^{\rm f}({\bf k},0)\,{\rm e}^{-kz}.
\label{e31}
\end{equation}

The boundary conditions (\ref{e23}) in the 2D
Fourier representation read
\begin{eqnarray}
{1\over 2}(ik_xu^{\rm f}_z-ku^{\rm f}_x)&=&-ik_xk\chi({\bf k},0),
\nonumber\\
{1\over 2}(ik_yu^{\rm f}_z-ku^{\rm f}_y)&=&-ik_yk\chi({\bf k},0),
\label{e32}\\
-k\left(K+{4\over 3}\mu\right)u^{\rm f}_z
&+&\left(K-{2\over 3}\mu\right)
\left(ik_xu^{\rm f}_x+ik_yu^{\rm f}_y\right)
\nonumber\\
&=&-p-2\mu k^2\chi({\bf k},0)+U({\bf k},0).
\nonumber
\end{eqnarray}
We have used equation (\ref{e31}) to evaluate the derivatives
on the left hand sides of (\ref{e23})  and the ansatz (\ref{e25})
together with relations (\ref{e28}-\ref{e30}) in the right
hand sides of (\ref{e23}).

From the first and second conditions of (\ref{e32}) 
one eliminates
$u^{\rm f}_x$ and $u^{\rm f}_y$ obtaining
$ik_xu^{\rm f}_x+ik_yu^{\rm f}_y=-ku^{\rm f}_z-2k^2\chi$.
Using this relation in the third condition of (\ref{e32}) 
one arrives at the $z$-component of the free displacement
\begin{equation}
u^{\rm f}_z={p-U({\bf k},0)-2k^2\left(K-{5\over 3}\mu\right)\chi({\bf k},0)
\over 2k\left(K+{1\over 3}\mu\right)}.
\label{e33}
\end{equation}

With $u^{\rm f}_z$ from (\ref{e33}) and the first and the second
equations of (\ref{e32}) one readily evaluates the parallel
displacements $u^{\rm f}_x$ and $u^{\rm f}_y$, respectively. Here
we want to concentrate on the component $u^{\rm f}_z$ perpendicular
to the surface.

\subsection{Surface corrugation}
The corrugation of the surface is given by the total
displacement in the $z$-direction $u_z=u^{\rm f}_z+u^{\rm i}_z$. Adding
formulas (\ref{e33}) and (\ref{e29}) we obtain
\be
u_z({\bf k})&=&{p-U({\bf k},0)+4k^2\mu\,\,\chi({\bf k},0)\over
2k\left(K+{1\over 3}\mu\right)}\nonumber\\
u_z({\bf r})&=&\sum_{\bf k}
{\rm e}^{i\bf k \bf r} u_z({\bf k}).
\label{e34}
\ee

Formula (\ref{e34}) is the final result of the general
part of our discussion. It provides the 2D Fourier
decomposition of atomic displacements at the surface
layer. In our notation a positive displacement $u_z$
corresponds to the atomic motion inwards the
superconductor.

\section{Numerical results for the electrostatic interaction}
To proceed we assume that the interaction between the
super-electrons and the ionic lattice is exclusively
carried by the mean electrostatic field. This approximation
neglects the effect of the lattice density on the electronic
band structure and on the phonon band structure and for
many materials it might lead to quantitatively incorrect
predictions. In this paper we adopt this approximation
for its simplicity.

\subsection{Surface dipole}
As mentioned in the introduction, in the electrostatic
approximation, the effective potential is proportional
to the electrostatic Bernoulli potential $\varphi$, i.e.,
\begin{equation}
U({\bf r},z)=\rho\varphi({\bf r},z).
\label{e35}
\end{equation}
The Bernoulli potential is a function of the GL
function\cite{LKMB01}
\begin{eqnarray} 
e\varphi
&=&-{\partial\alpha\over\partial n}|\psi|^2-
{1\over 2}{\partial\beta\over\partial n}|\psi|^4,
\label{chi6}
\end{eqnarray}
where $\alpha=\gamma(T^2-T_{\rm c}^2)/2n$ and
$\beta=\gamma T^2/2n^2$ are the GL parameters
expressed in terms of Sommerfeld's gamma. These values follow
from the Gorter-Casimir two-fluid model in the limit
$T\to T_{\rm c}$, see Ref.~\onlinecite{LKMBY07}.
One can also use the limiting BCS values of $\alpha$
and $\beta$ which are larger by factors $1.43$ and
$2.86$ than the two-fluid limit, respectively.\cite{LMKBS08}

The pressure at the surface is caused by the surface
dipole\cite{LMKBS08}
\begin{equation}
p({\bf r})=\rho\varphi_+({\bf r})-
\rho\varphi_-({\bf r}),
\label{e36}
\end{equation}
where $\varphi_\pm({\bf r})=\varphi({\bf r},\pm\epsilon)$
with $\epsilon>0$ being an infinitesimal distance.
In reality, $\epsilon$ has to exceed the Thomas-Fermi screening
length and the BCS coherence length. The potential $\varphi_+$
is the extrapolation of the internal potential towards the
surface, and $\varphi_-$ represents the potential out of the crystal.

The surface value of the effective potential is defined
as a limit from inside, therefore
\begin{equation}
U({\bf r},0)\equiv \lim\limits_{\epsilon\to0}U({\bf r},\epsilon)=
\rho\varphi_+({\bf r}).
\label{e37}
\end{equation}
The displacement of the surface atoms (\ref{e34}) for the
electrostatic approximation reads
\begin{equation}
u_z=-{\rho\over 2k\left(K+{1\over 3}\mu\right)}
\varphi_-({\bf k})+
{2k\mu\over K+{1\over 3}\mu}\chi({\bf k},0).
\label{e38}
\end{equation}

As one can see, the potential $\varphi_+$ cancels and we are left
with the potential $\varphi_-$. According to the Budd-Vannimenus
theorem for superconductors, the potential above the surface is
proportional to the density of the free energy,\cite{LMKMBS04,LMKMBSb04}
\begin{eqnarray} 
e\varphi_-
&=&-{\alpha\over n}|\psi|^2-
{1\over 2}{\beta\over n}|\psi|^4,
\label{chi8}
\end{eqnarray}
which one conveniently evaluates within the Ginzburg-Landau theory.
For details of the Bernoulli potential and the surface dipole see
textbook~[\onlinecite{LKMBY07}].

\subsection{Bulk forces}

The internal value of the Bernoulli potential contributes
only via the subsidiary function $\chi$. In general one
needs the complete 3D solution of the vortex lattice near
the surface. Numerical studies have shown that one can neglect
the $z$ dependence of the amplitude of the GL function everywhere
including the close vicinity of the surface.\cite{B97c} The
amplitude variation is about 1\%. Since the Bernoulli potential
depends exclusively on the amplitude of the GL function, 
the potential has the same value
as deep in the bulk,
$\varphi({\bf r},z)\approx\varphi_\infty({\bf r})$, for any $z$.

Neglecting the $z$ dependence of the potential, the integral in
(\ref{e28}) becomes trivial giving
\begin{equation}
\chi({\bf k},0)=
{1\over 2k^2}{\rho\over K+{4\over 3}\mu}
\varphi_\infty({\bf k}).
\label{e40}
\end{equation}
The potential $\varphi_\infty$ is conveniently found from
(\ref{chi6}) using the GL function deep in the sample. For a fixed
magnetic field, the GL function deep in the sample is the same as in
the infinite sample, therefore it can be evaluated as an effectively
two-dimensional problem.

\subsection{Surface corrugation}
The displacement of the surface atoms results from
(\ref{e38}) and (\ref{e40}) as
\begin{equation}
u_z=-{\rho\over 2k\left (K+{1\over 3}\mu\right)}\left [
\varphi_-({\bf k})-
{2\mu\over K+{4\over 3}\mu}
\varphi_\infty({\bf k})\right].
\label{e41}
\end{equation}
One can see from (\ref{e41}) that the relative contribution of the
bulk and the surface potentials depends on the shear modulus $\mu$.
For soft materials with $\mu\ll K$
the
bulk potential gives a negligible contribution so that the surface
corrugation is driven only by the electrostatic potential above
the surface. In this case the corrugation is independent of the
density derivatives of the GL parameters.

\begin{table}
\begin{tabular}{|c|c|c|c|c|c|c|c|}
\hline
&${\gamma T_{\rm c}^2\over 4 n}$ & $\kappa$& n & $\partial \ln T_{\rm c}\over\partial \ln n$
& $\partial \ln \gamma\over\partial \ln n$ & $ E$ & $\sigma$
\cr
&[$\mu$eV]&&[$10^{28}$m$^{-3}$]& &&[GPa]&
\cr
\hline
Nb& $4.585$ &1.5&2.2& 0.74\cite{LKMBY07} &0.42\cite{LKMBY07}&
105\cite{note}&0.4\cite{note}
\cr
YBCO & 750 & 65&0.5&-4.82\cite{note2} &-4.13\cite{note2}&
200\cite{SVLA04,KNSS91} 
&0.2\cite{KNSS91} 
\cr
\hline
\end{tabular}
\caption{Material parameters of Niobium and YBa$_2$Cu$_3$O$_7$:
condensation energy per particle, GL parameter, particle density
and logarithmic derivatives of the critical temperature and the
linear coefficient of the specific heat $\gamma$, Young's modulus and Poisson ratio.}
\label{tab}
\end{table}

Many authors dealing with deformable superconductors
prefer to express elastic coefficients via the Poisson
ratio $\sigma=\left(K-{2\over 3}\mu\right)/
\left(2K+{2\over 3}\mu\right)$ and the Young modulus
$E=3K(1-2\sigma)$. In this notation the displacement of surface
atoms (\ref{e41}) reads
\begin{align}
u_z={(1-2\sigma)(1+\sigma)\over k\, E (1-\sigma)}
\left \{ (1-\sigma) p({\bf k}) 
-\rho \sigma \varphi_\infty({\bf k})\right \}.
\label{e44}
\end{align}
Deriving (\ref{e44}) we have used that the Bernoulli potential is 
nearly independent of $z$ so that $\varphi_+=\varphi_\infty$. 
The pressure at the surface (\ref{e36}) thus equals $p=
\rho(\varphi_\infty-\varphi_-)$.

\begin{figure}[t]
\psfig{file=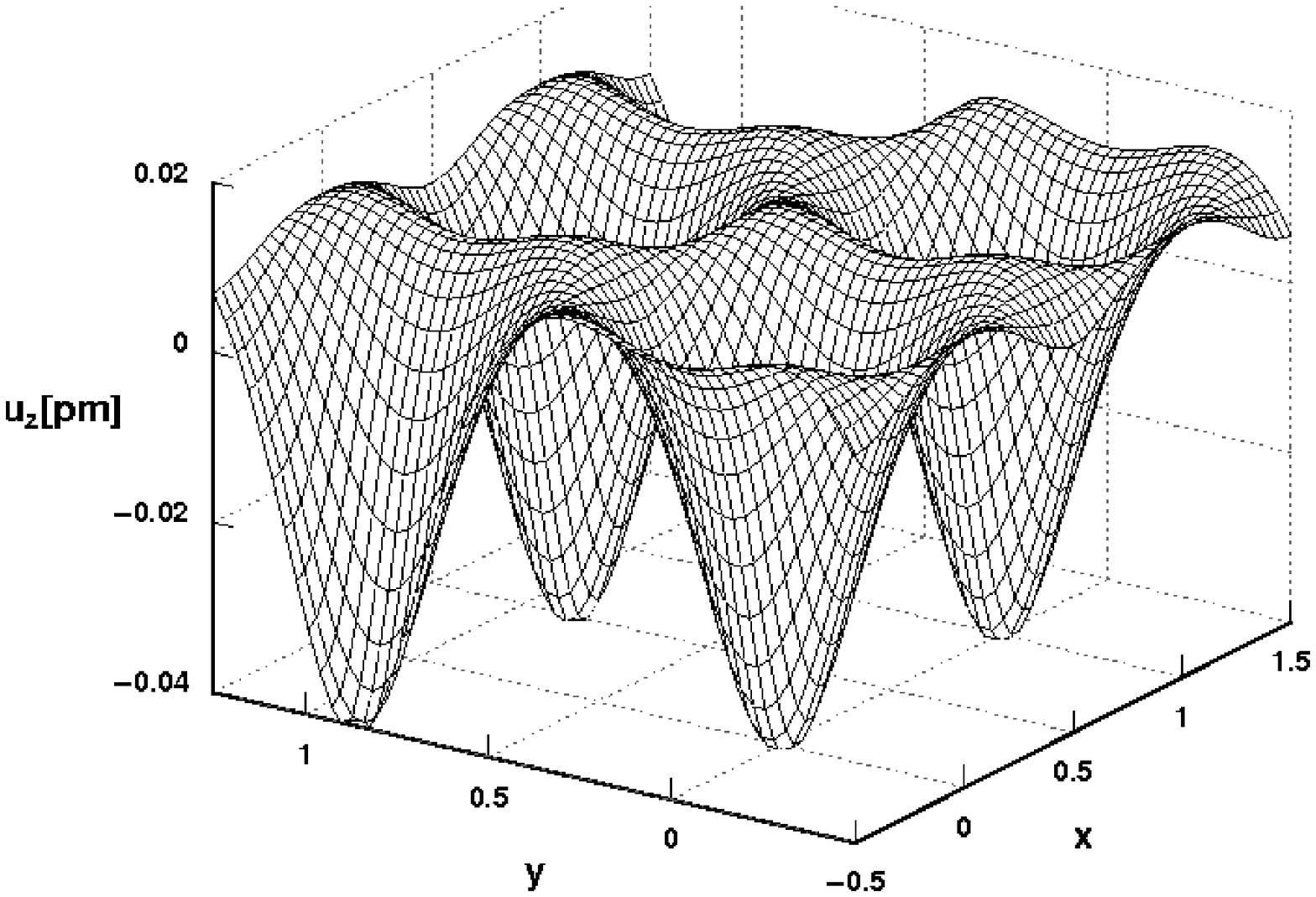,width=8cm,angle=0}
\vskip3mm
\psfig{file=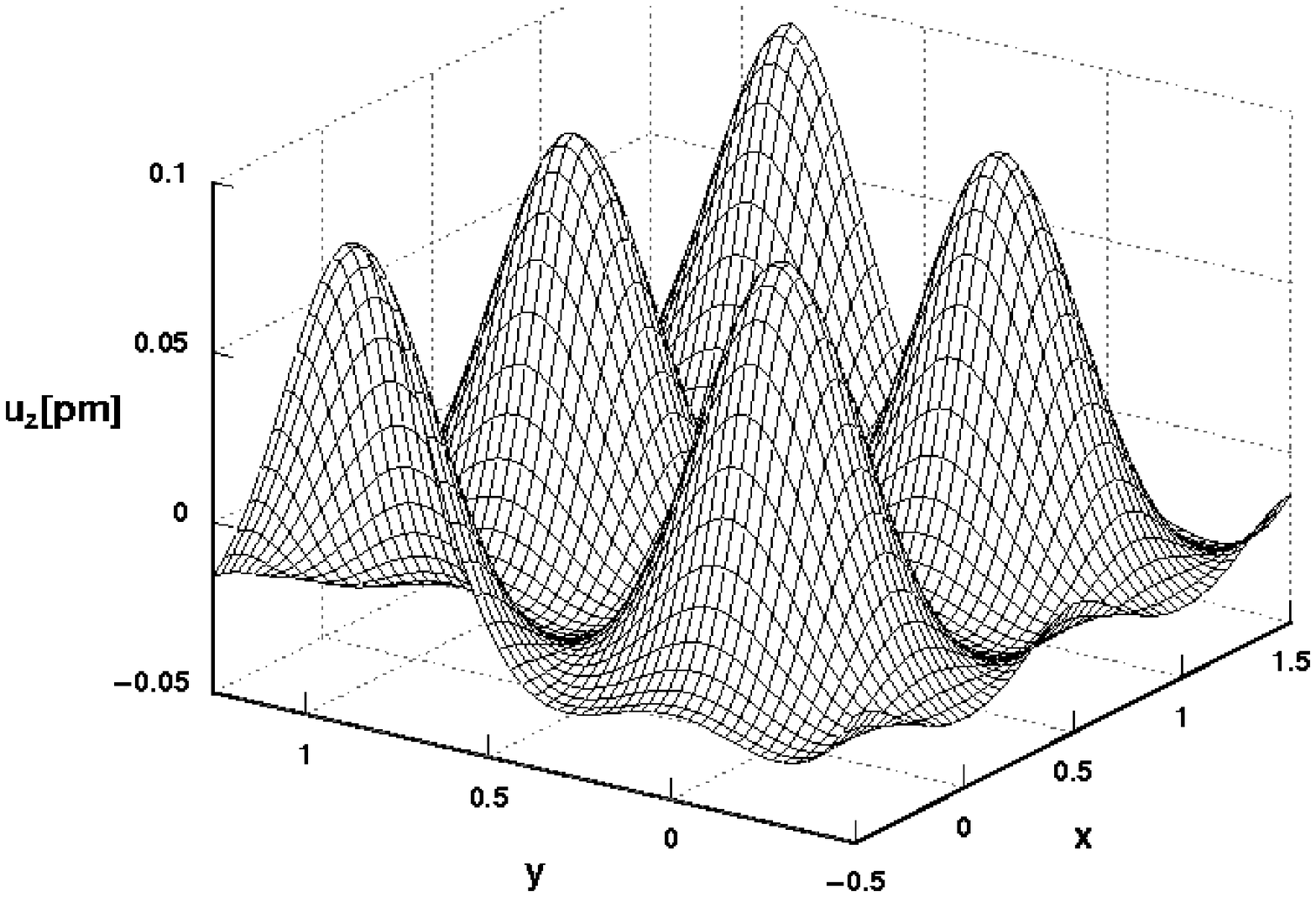,width=8cm,angle=0}
\vskip3mm
\psfig{file=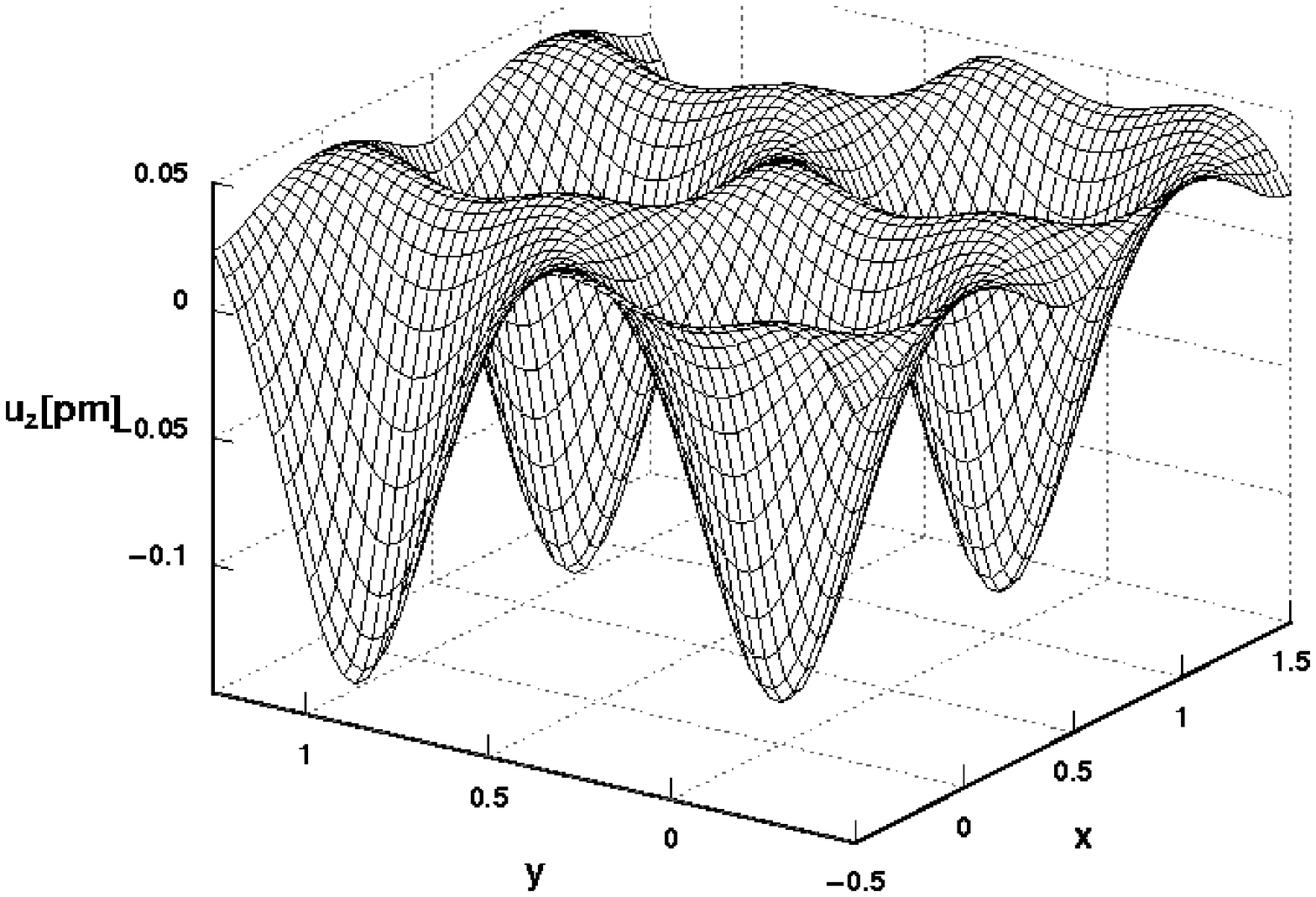,width=8cm,angle=0}
\caption{Surface corrugation $u_z({\bf r})$ of Nb (above)
compared with the value neglecting surface dipoles (middle)
and only surface dipoles (below). The temperature and the
mean magnetic field are
$T=0.95~T_{\rm c}=9$~K and $\bar B=0.21~B_{\rm c2}(T)=6.4$~mT.
Length unit is the vortex distance $a=128$~nm. 
\label{niobium}}
\end{figure}

\begin{figure}[h]
\psfig{file=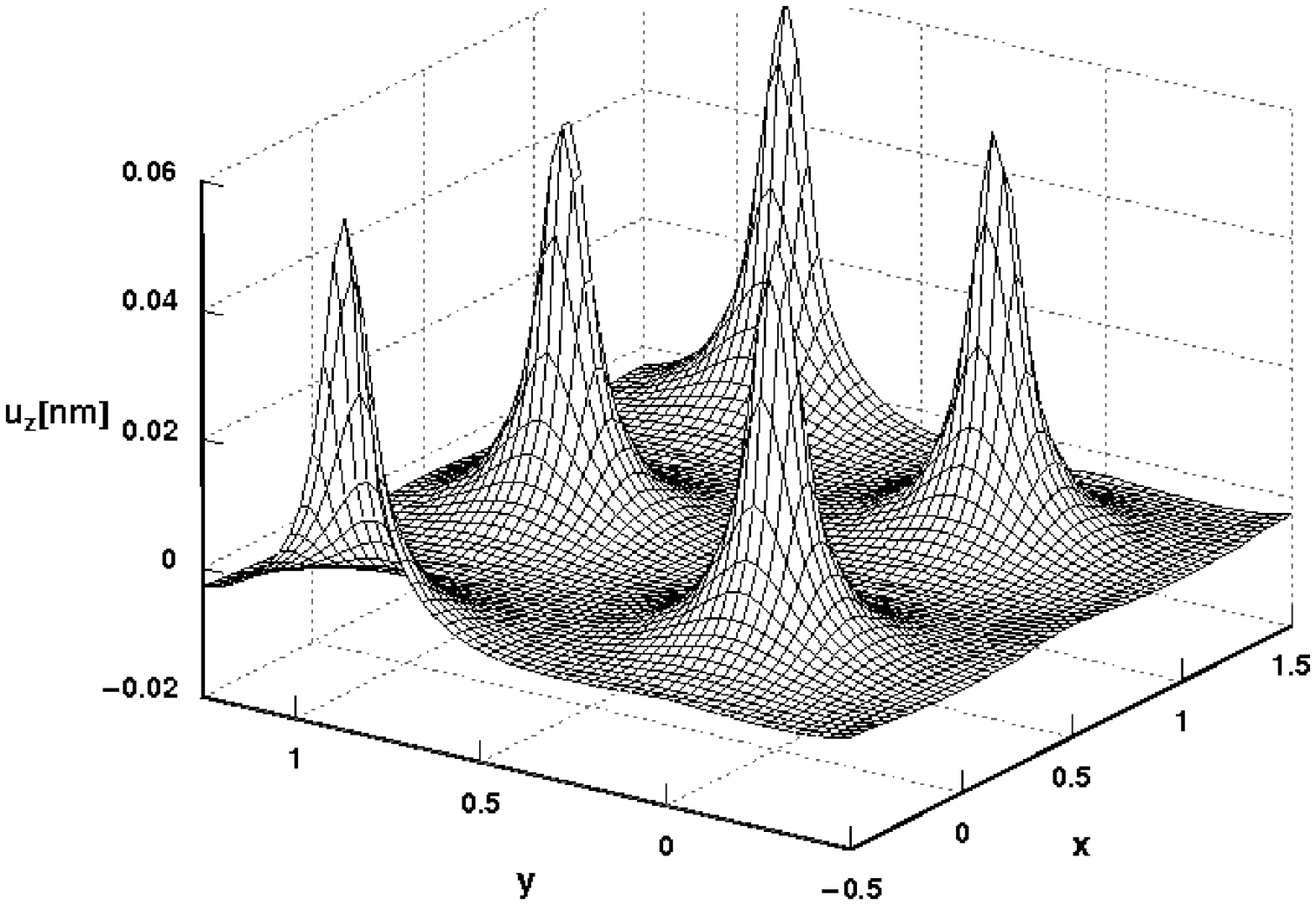,width=8cm,angle=0}
\vskip3mm
\psfig{file=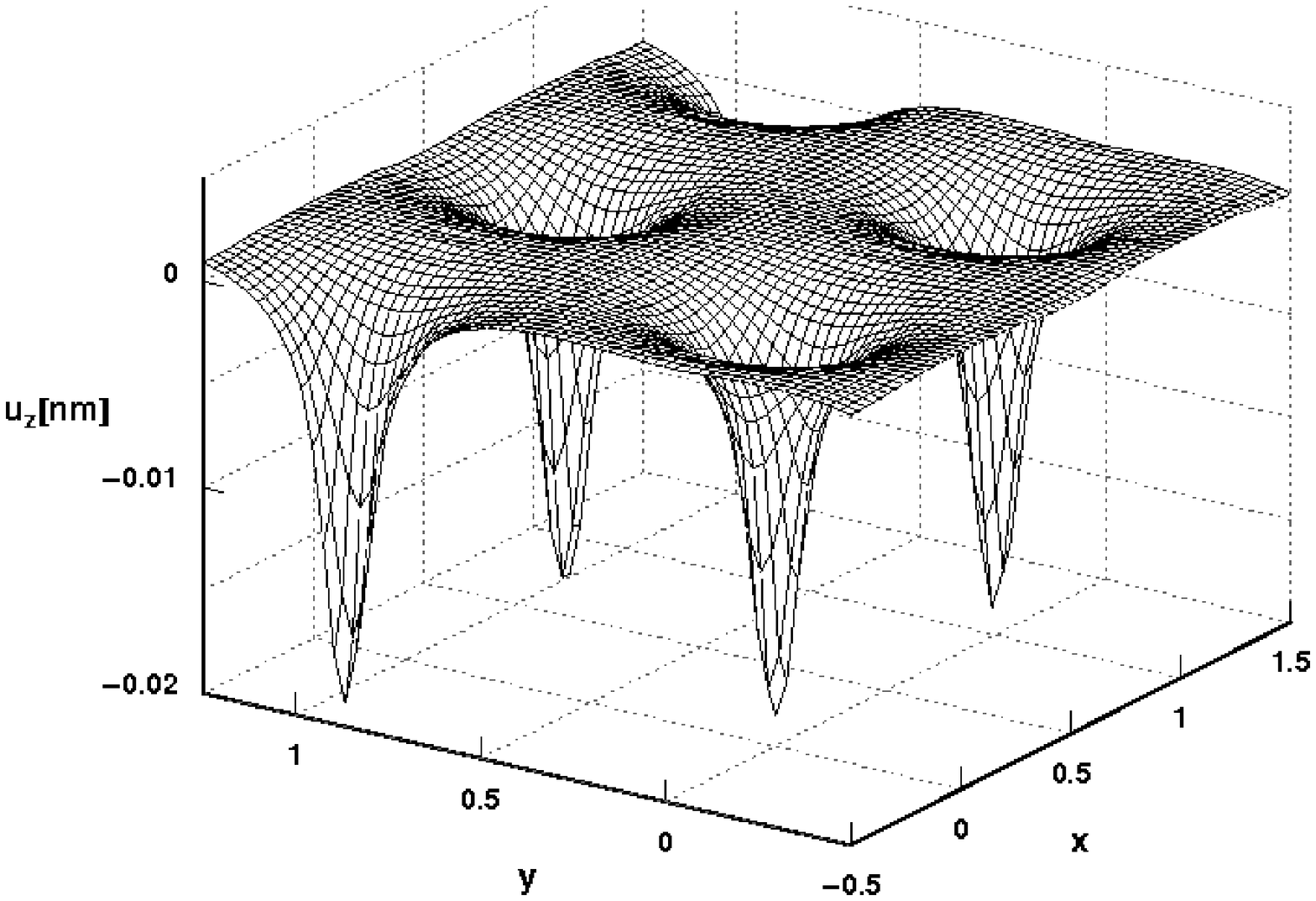,width=8cm,angle=0}
\vskip3mm
\psfig{file=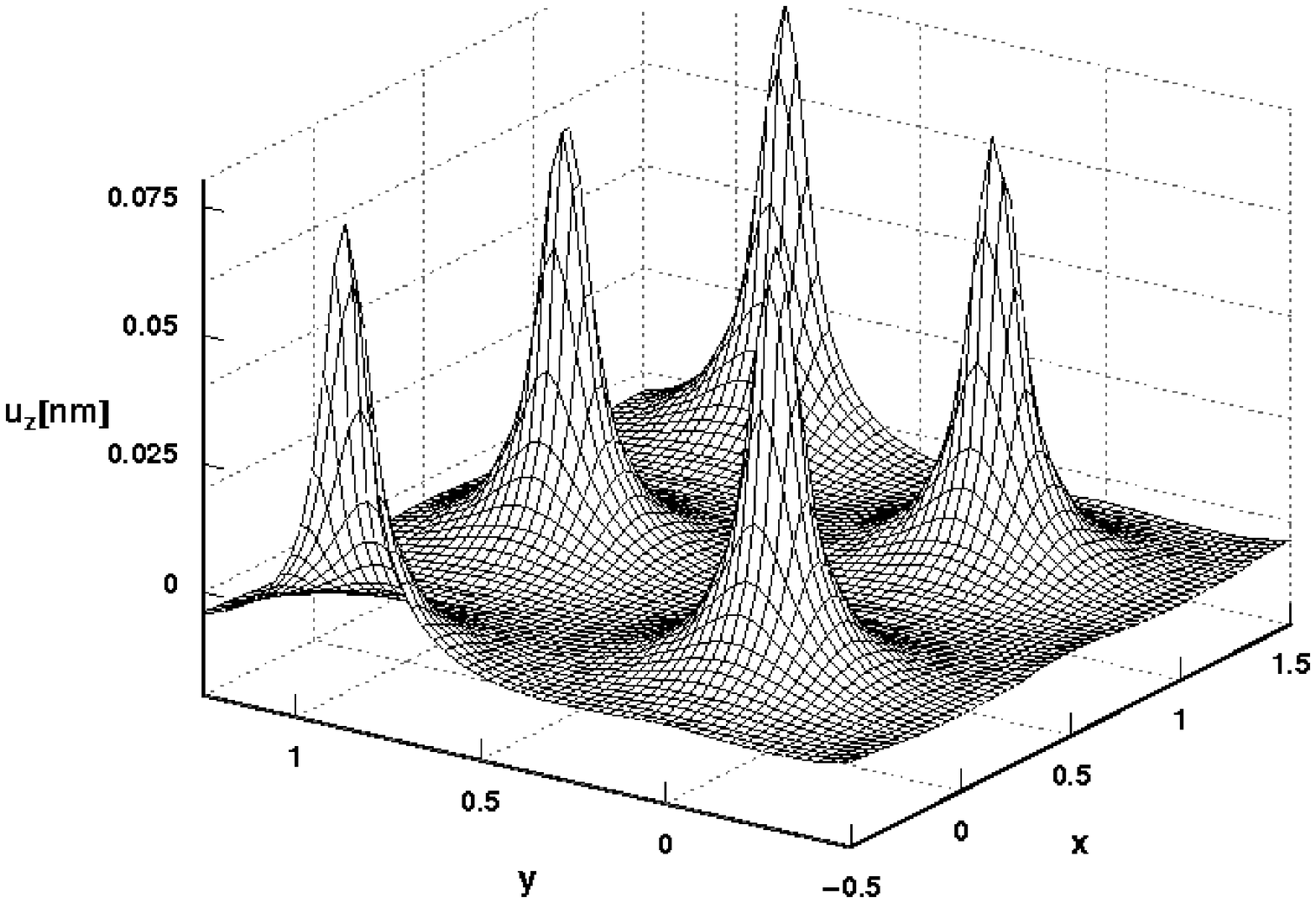,width=8cm,angle=0}
\caption{Surface corrugation $u_z({\bf r})$ of YBa$_2$Cu$_3$O$_7$ (above)
compared with the value neglecting surface dipoles (middle)
and only surface dipoles (below). The temperature and the mean magnetic
field are
$T=0.67~T_{\rm c}=60$~K and $\bar B=0.01~B_{\rm c2}(T)=0.6$~T.
Length unit is the vortex distance $a=58$~nm. 
\label{YBCO}}
\end{figure}

The numerical study of the surface corrugation (\ref{e44}) is 
performed with the parameters listed in the table~\ref{tab}. 
Figure~\ref{niobium} presents the surface corrugation of Niobium
with the GL parameters $\kappa$ increased by impurities to $1.5$.
The scale is in pm=$10^{-2}$~\AA, therefore the maximum deformation
$\sim 5~10^{-4}$~\AA{} is far too small to be detected by recent
scanning microscopes. In our convention the negative $u_{\rm z}$
corresponds to atoms displaced out of the crystal.

The total deviation of atoms at the surface is seen in the upper 
part of figure~\ref{niobium}. The middle part shows the deviation 
evaluated omitting the surface dipole, i.e., setting $p=0$ in 
formula (\ref{e44}). As one can see, such approximation for 
Niobium leads to the opposite sign of the atomic displacement. 
The lower part shows the surface deformation evaluated from the 
surface dipole only. This approximation overestimates the amplitude 
more than twice.

The magnitude of the surface corrugation is directly proportional
to the condensation energy per particle, which for Niobium has the 
value ${\gamma T_{\rm c}^2\over 4 n}=4.6 \,{\rm \mu eV}$. In 
conventional superconductors this condensation energy per particle 
is small because of the small critical temperature, $T_{\rm c}=9.5$~K 
and the large density of particles, $n=2.2~10^{28}$/m$^3$.

In high-$T_{\rm c}$ superconductors the critical temperature
is larger by a factor of ten,  while the density of particles is 
lower by a factor of ten. This leads to an appreciably larger 
condensation energy per particle. For example, 
in YBa$_2$Cu$_3$O$_7$ one has $T_{\rm c}=90$~K and 
$n=5\times 10^{27}$~m$^{-3}$ resulting into
${\gamma T_{\rm c}^2\over 4 n}=750 \,{\rm \mu eV}$. 
Figure~\ref{YBCO} presents a corrugation of the YBa$_2$Cu$_3$O$_7$ 
surface with the surface parallel to the $a$-$b$ planes,
i.e., the magnetic field along the $c$ axis. As one can see, the
maximum displacement of the surface atom is more than 0.5~\AA ,
which can be observed by the scanning microscope. Again, neglecting
the surface dipole one arrives at the opposite displacements, while
using only the surface dipole one overestimates the magnitude.

Perhaps we should note that application of the isotropic model to
the layered structure of YBa$_2$Cu$_3$O$_7$ is not justified.
One clearly needs more elastic coefficients to describe the 
deformation of this highly anisotropic material.\cite{SO91} The 
presented data can thus serve only as an order of magnitude estimate.

\section{Conclusions}

Magnetic field entering the superconductor in the \mbox{form} of vortices
induces a corrugation of the surface. In conventional superconductors
the displacement of surface atoms is of the order of $10^{-4}$~\AA,
which is too small to be observed by recent experimental tools.
In high-$T_{\rm c}$ superconductors one can expect amplitudes
$\sim 10^{-1}$~\AA\ which is in the reach of scanning force
microscopes.

Our results for Niobium and YBa$_2$Cu$_3$O$_7$ show that \mbox{among}
the forces that drive the surface corrugation the dominant one is due to the
surface dipole. The contribution of the bulk potential to the surface
corrugation is opposite to the contribution of the surface, therefore
it reduces the magnitude of the atomic displacement.

Since the deformation of the crystal near the surface differs from the
deformation in the bulk, one can \mbox{expect} that the surface terms play
an important role in the {\v S}im{\'a}nek contribution to the vortex
mass in thin \mbox{layers}. We leave this problem for a future work.

\medskip
This work was supported by research plans
MSM 0021620834 and No. AVOZ10100521, by grants
GA\v{C}R 202/07/0597, 202/08/0326 and GAAV 100100712, by PPP project of 
DAAD, by  
DFG Priority Program 1157 via GE1202/06 and the BMBF and 
by European ESF program NES.

\bibliography{bose,delay2,delay3,gdr,genn,chaos,kmsr,kmsr1,kmsr2,kmsr3,kmsr4,kmsr5,kmsr6,kmsr7,micha,refer,sem1,sem2,sem3,short,spin,spin1,solid,deform}

\begin{thebibliography}{26}
\bibitem[{not({\natexlab{b}})}]{note2}
\bibinfo{note}{The pair-able charge per Cu atoms is $-0.4335\,e$. In Fig.~2.16
  Plakida\cite{Plak95} shows results of Cava\cite{CHHBMRKPR90} according to
  which the charge transfer $-0.03\,e$ from chains to planes per Cu site leads
  to a decrease of the critical temperature by 30~K. This corresponds to
  ${\partial\ln T_{\rm c}\over \partial\ln n}=-4.82$. From Fig.~3 of
  [\onlinecite{LMCL93}] we can see that the specific heat coefficient drops at
  the same time from $4.4$~mJ/gK$^2$ to $3.0$~mJ/gK$^2$ which gives
  ${\partial\ln \gamma\over \partial\ln n}=-4.13$. This is only a rough
  estimate since the specific heat data include chains while we need the change
  in plane only.}
\end{thebibliography}

\end{document}